\documentclass[structabstract]{aa}
\usepackage{graphicx}
\usepackage{rotating}
\usepackage{txfonts}
\usepackage{natbib}   % biblio par bibtex, avec natbib.sty fourni dans le package AaA

\begin{document}

\title{Towards a new full-sky list of radial velocity standard stars}
\author{F. Crifo
       \inst{1}
\and 
G. Jasniewicz
\inst{2}
\and 
C. Soubiran
\inst{3}
\and 
 D. Katz
 \inst{1}
\and 
A. Siebert
\inst{4}
 \and 
L. Veltz
 \inst{5, 7}
\and 
S. Udry
\inst{6}
}

\institute{
GEPI, Observatoire de Paris, CNRS, Universit\'e Paris Diderot; 
 5 Place Jules Janssen, 92190 MEUDON, France
          \and
UMR CNRS/UM2 GRAAL, CC 72,  
Universit\'e Montpellier2, 34095 MONTPELLIER Cedex 05, France
\and
Universit\'e Bordeaux 1, CNRS, LAB, F-33270 
FLOIRAC, France
\and
Observatoire astronomique, Universit\'e de Strasbourg, 
CNRS, 11 rue de l'Universit\'e, 67000 STRASBOURG, France
\and
Astrophysikalisches Institut Potsdam, POTSDAM, Germany
\and
Observatoire de Gen\`eve, 51 Ch. des Maillettes, 
1290 SAUVERNY, Switzerland
\and
Institut d'Astrophysique Spatiale, Universit\'e Paris-Sud, 
F-91405 ORSAY cedex, France
}

% \date{Received September 15, 1996; accepted March 16, 1997}

 \abstract
  % context heading (optional)
   {}  %leave it empty if necessary 
   % aims heading (mandatory)
{The calibration of the Radial Velocity Spectrometer (RVS) onboard the
ESA Gaia satellite (to be launched in 2012) requires a list of
standard stars with a radial velocity (RV) known with an accuracy of at least 300\,m s$^{-1}$.
The IAU Commission 30 lists of RV standard stars are too bright and not dense enough.
}
 % methods heading (mandatory)
{We describe the selection criteria due to the RVS constraints for building an 
adequate full-sky list of at least 1000 RV standards from catalogues already 
published in the literature.
}
 % results heading (mandatory)
{A preliminary list of 1420 candidate standard stars is built and its
properties are shown.
An important re-observation programme has been set up in order to insure
within it the selection of objects with a good stability until the end 
of the Gaia mission (around 2018). 
%\textbf{The present list of candidate standards} is available at CDS and usable for many other projects.
}
% conclusions heading (optional)
{The present list of candidate standards is available at CDS and usable for many other projects.
}


\maketitle

%________________________________________________________________

\section{Introduction}

Since the first catalogue of stellar radial velocities published by \citet{Moore},
the number of measured stellar radial velocities (RV) has increased 
considerably, thanks to the very wide uses of these data, from galactic
dynamics to stellar atmospheric motions and the discovery of extra-solar planets.
The General Catalogue of stellar Radial Velocity \citep[GCRV, see][]{GCRV} 
has been used for many different purposes.
Today, very extensive RV surveys are in progress, like 
the Radial Velocity Experiment \citep[RAVE 
project, see][]{RAVE},
%  Zwitter et al 2008, 
or the SDSS-SEGUE survey 
 \citep [see][]{SEGUE};
or in preparation, like the Gaia-RVS \citep[see][]{Katz_SF2A}, and the LAMOST 
survey \citep[see][]{2009AAS...21341614N}. 
Each of these surveys has 
its own scientific goals, but all need a wavelength calibration
 which guarantees the quality of spectroscopic data,
particularly radial velocities (RV) and rotational velocities.
These calibrations rely on the radial velocity
zero-point (RVZP) of the instruments, which can evolve with time during the 
life of the project. 
A way to take
care of these RVZP is the continuous observation of "standard" 
stars and/or "reference" objects such as asteroids. 
The concept of ``standard star'' and ``set of standards'' has been carefully examined by 
Batten \citep{Battena,Battenb} and is based on the physical notion 
of ``stability'';
this notion of ``stability'' is of course limited by other physical
concepts \citep[for the fundamental definition
of radial velocity, see][]{Lindegren} and is very dependent
on the accuracy  expected for the project. It is the reason why each project, 
whether ground-based or space-based, has to establish an appropriate list 
of such "standard stars`` .

In this paper we discuss the construction of a new full-sky list of candidate
RV standard stars developed specifically within and for the Gaia project,
 but which will hopefully be very useful for many other astronomical projects.
Indeed, this list of rather bright stars is established for the RVS (Radial Velocity
Spectrometer) survey on board the Gaia satellite \citep[see][]{Gaia}, which will use it
in close interaction with asteroids for its RVZP . The strategy
for building this list is somewhat different from that adopted 
by the Space Interferometry Mission
\citep[SIM project, see][]{SIM}, which plans to build a reference astrometric grid
over the whole sky by means of radio-loud quasars and
a ''Grid Giant Star Survey'' composed of RV-stable red K giant stars (Bizyaev et al. 2006).
While the large ground-based RV-surveys like SDSS-SEGUE, RAVE or the ELODIE Archive
\citep[see][]{ELODIE} use calibration
lamps permanently installed on the instrument to determine the wavelength scale,
standard stars also allow to check the good status of the spectrograph such as stability 
or possible slow drifts.

The process of building this list in addition to the official IAU Commission 30 lists
 (available at: \textbf{http://obswww.unige.ch/$\sim$udry/std/std.html}) may be 
compared to the similar but much older process in astrometry:
a short basic list of very bright stars with very accurate positions, like the FK
series of astrometric catalogues;
supplemented by larger and denser lists of fainter objects, like the AGK catalogues.

We discuss first the former lists of RV standard stars and the 
available data for building a new larger one.
Then we describe in detail the constraints and criteria adopted, 
 the way we have selected the candidate stars, and the properties
of the resulting list.  
Finally we develop the on-going strategy for re-observing 
our sample in both hemispheres.

\section{The IAU Commission 30 lists of RV standards}

\begin{figure}[htp]

\begin{center}
%\resizebox{\hsize}{!}{\includegraphics[angle=-90]  {figures/bv_v_iaustd.ps}}
%\resizebox{\hsize}{!}{\includegraphics[angle=-90]  {figures/carteIaustd.ps}}
\resizebox{\hsize}{!}{\includegraphics[angle=-90]  {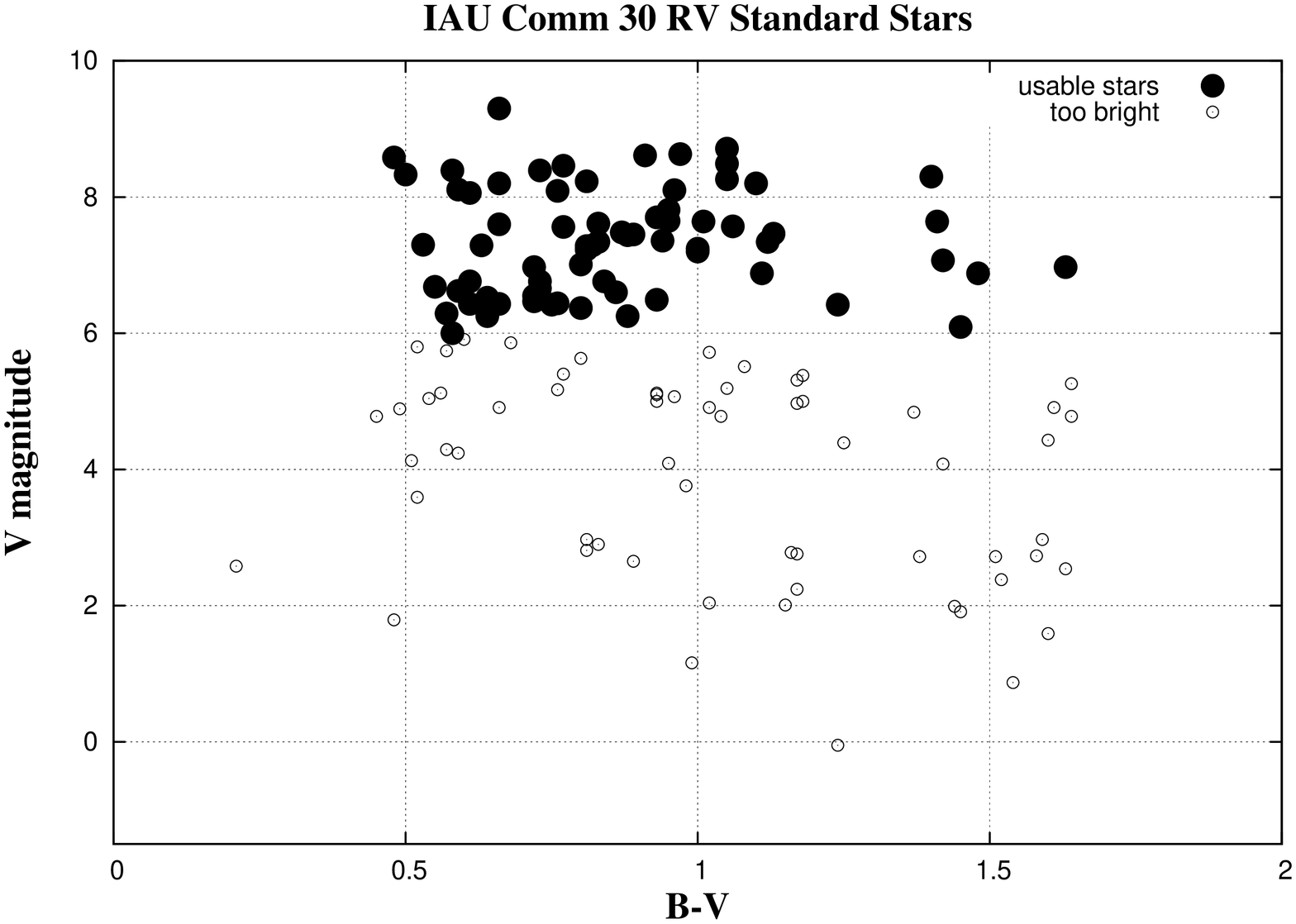}}
\resizebox{\hsize}{!}{\includegraphics[angle=-90]  {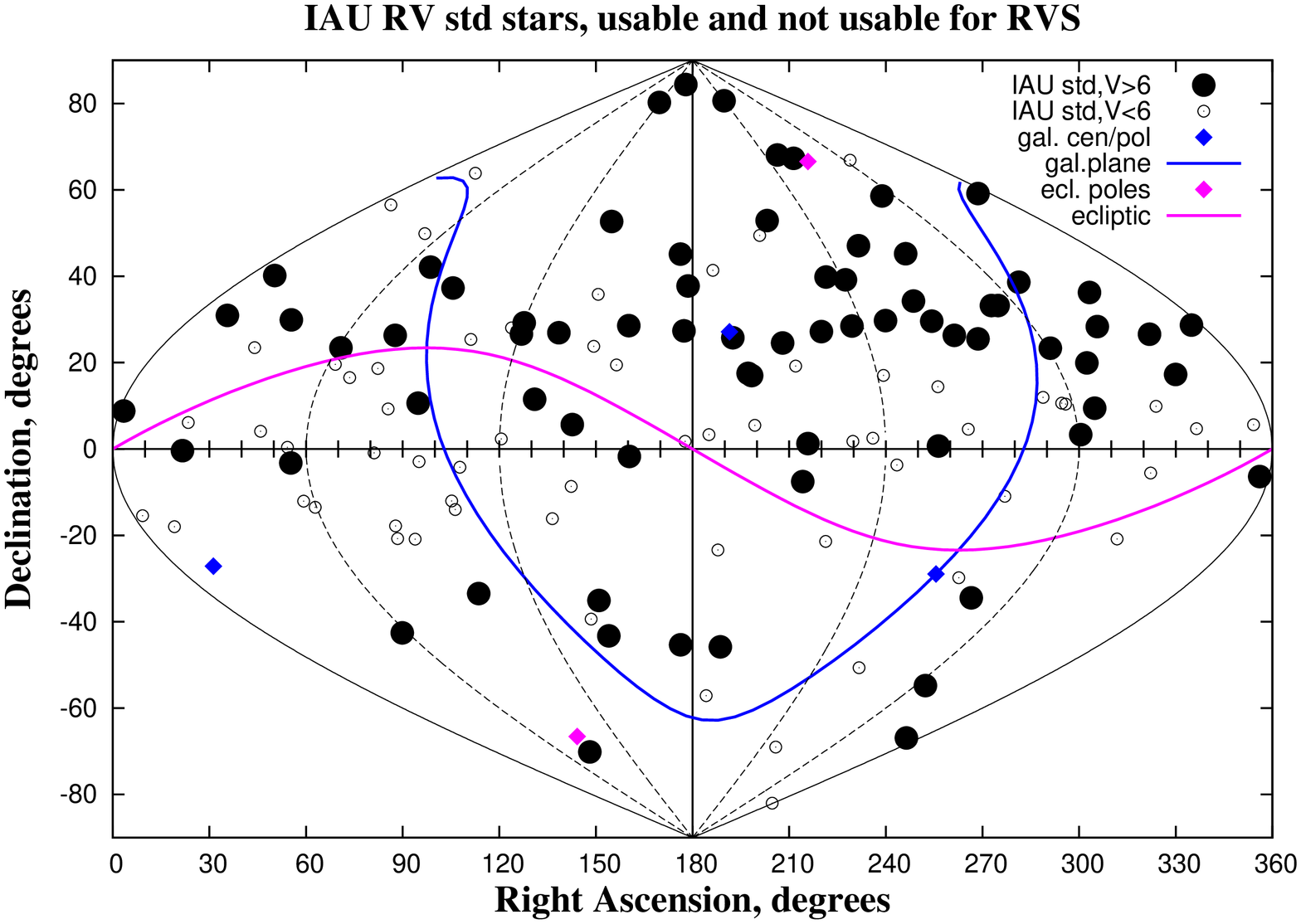}}

 \caption{
Properties of the IAU commission 30 RV standard stars.
Top: distribution in the colour-magnitude plane. 
Stars brighter than V=6  (small open circles) are too bright for RVS 
standards, and will therefore not be kept for our list.
Filled circles indicate IAU standards fainter than V=6, suitable for RVS.
Bottom: sky distribution of the IAU standards (symbols as for the top panel). 
A large fraction of the usable standards is
located in a strip around $\delta=+30\degr$, with a clear lack of 
standards in the southern hemisphere. }
 \label{iau-std}
\end{center}
\end{figure}

IAU Commission 30 has set up a working group devoted to RV standards.
Two lists of standards are available at the above address:
\begin{itemize}
 \item CORAVEL standard stars, with only a limited accuracy ( on the order of 300\,m s$^{-1}$; 
some much worse); 107 objects;
\item "New" ELODIE-CORAVEL high-precision standard stars: 37 objects, with $0.57 \leq B-V \leq
0.94$, 
and an accuracy of about 50\,m s$^{-1}$.
\end{itemize}
These objects have been repeatedly observed for years, and will constitute a firm
 basis for a new list. 
However they are not numerous enough for the RVS: only 144 objects, most of them too bright 
for the RVS (see Fig. \ref{iau-std}) .

It is therefore necessary to search for more ``standards'', which have of course
a shorter observational history.

We note that several former
"stable" stars in these lists recently revealed to be variable with an amplitude about
a hundred of m s$^{-1}$  thanks to the use of very high-precision spectrometers
such as CORALIE \citep[see][]{CORALIE} and HARPS \citep[see][]{HARPS}, devoted to the 
search of extrasolar planets.

% \subsection{Other lists and catalogues}

\section{Constraints due to the Gaia-RVS}

The Gaia RVS is an integral-field spectrograph with \textit{no} calibration 
device onboard owing to payload size and weight constraints.
Therefore calibration sources must be found among the many programme 
objects and used as references. For the RVS, the best objects are
the asteroids; unfortunately only very few are bright enough, and their 
distribution across the sky is not uniform because they are concentrated in the 
vicinity of the ecliptic. Although they are the ultimate
references (their RV can be accurately calculated from celestial mechanics),
they must be supplemented by a list of suitable stars. 
Descriptions of the RVS may be found in \citet{Gaia},
and in \citet{Katz_SF2A} for an up-to-date description.

In this paper, we present the construction of the list of 
possible standards for the RVS instrument. Our selection, based 
on available ground-based data, is performed according to criteria 
tuned for the Gaia-RVS specific needs, like the RVS
geometry, its feeding and read-out system as well as the Gaia 
scanning law, leading to specific constraints. These
requirements will just be listed for a good unterstanding of the 
procedure; but they will not be established nor developed here. 
The resulting list can be used totally independently
of the RVS for many different purposes, because it covers the whole sky.

\subsection{Magnitude range}
Very bright stars are not taken into account for RVS instrumental 
configuration; and faint stars with a low S/N have
a low accuracy and cannot be used as references.
The RVS spectral interval 847--874 nm is contained in the $I$ band. 
The magnitude over this small interval is denoted $G_{RVS}$, and can 
be expressed as a polynomial
transformation equation between  the usual $V$ magnitude and the 
Cousins' $I$, or $I_C$ magnitude.
The magnitude limitations are of course given primarily in $G_{RVS}$.
The magnitude range for the RVS standard stars is defined as:
 
bright end: $V = 6.0$ (on-board limitation);  

faint end: $G_{RVS} = 10.0$, corresponding to $V\sim11$ for K1III giants.

It can be seen in Fig. \ref{iau-std} that many IAU standards are therefore not usable
because they are too bright.

\subsection{Size of the list}

Some 1000 standard stars are needed for a best performance of the RVS 
reduction algorithm \citep{Guerrier}. 
%As this number is proper to RVS, it is not justified here.
However, this is the number of stars aimed at in the final list: given 
that some of the presently selected stars will not satisfy 
all the constraints, in particular those concerning binarity and variability,
it was decided to extend the size of the candidate standard stars to 
some 1500 objects. 
All these stars are re-observed before and during the mission; 
some will be dropped from the present list, 
and the final list to be kept will be known only at the 
\textit{end} of the Gaia mission, around 2018.
We nevertheless expect to have a good first selection at Gaia launch time 
(around the end of 2012).

\subsection{Initial accuracy in radial velocity}

The final accuracy on RVS velocities is expected to be of the order of 1\,km s$^{-1}$
for the brightest stars.
Therefore standard stars must be known \textit{a priori} with a better accuracy,
and the initial limit for selection was set to 300\,m s$^{-1}$; 
reducing it to 100\,m s$^{-1}$ would have been a criterion too
hard to fit, as less than the 1500 stars required were known with such an accuracy
when the search was begun. The remeasurement programme undertaken should
allow us to have at Gaia launch time a precision of the order of 
100\,m s$^{-1}$ for the whole list.

\subsection{Spectral range and spectral type range used}

The RVS spectral range is 847--874 nm; the choice of this small and up 
to now unusual range for RV is described in \citet{Munari}; 
see also \citet{Gaia}; it covers the three lines of the IR CaII triplet, 
which are visible in most stellar types, roughly B8 to M5V.
Moreover, in late-type stars it contains many narrow metallic lines, 
which are extremely useful for a good fit of the dispersion all
along the range; and in early-type stars several Paschen lines are visible.
Numerical simulations performed with IRAF over synthetic spectra show 
that the RV accuracy is reduced for stars earlier than F5.
Notice that if K-type red giants are very ``good'' stars with sharp lines, it
was found by  \citet{Bizyaev} during their preparatory study for the SIM 
mission that the best stable giants should satisfy $0.59 <(J-Ks) <0.73$
(roughly G7 to K2 giants). 
We will keep this constraint, converted into our colour indices, in order
to eliminate possible variables (see below).

\subsection{Photometric variability}

Variable stars are a real problem, irrespective of the cause of the variability: 
motions in the atmosphere, or close companions. In all cases, the variability
may be a reason for ``unstable'' radial velocity. Therefore all objects with 
an indication of variability in the Hipparcos Catalogue  \citep{ESA97} have been discarded, 
because we need objects stable until the end of the mission  (2018).

\subsection{Multiplicity}

Double stars must be eliminated as much as possible; all stars with some
indication of multiplicity have been deleted, even those with a long period,
because stability of RV is required until the end of the mission.
The RV lists and catalogues used have been carefully searched for indications of 
multiplicity. However some double stars are certainly still contained in our list.
The ongoing re-observations should allow us to eliminate most of them.

\subsection{Neighbourhood}

Because the RVS is a slitless spectrograph, somewhat comparable with the former
objective prism spectrographs, the spectra of the different stars may 
overlap in the focal plane, depending on the angular distance of the objects 
and the orientation of the dispersion. The overlap may disturb the main 
spectrum, inducing an error in the derived RV.
Because each star will be observed around 40 times during the mission, with 
various orientations for the scanning direction, we discarded stars 
with bright neighbours ($\Delta I <4$ ) within a circle of radius just above
the size of a spectrum on the detector ($\rho< 80\arcsec$).
This condition led to the deletion of many otherwise ``good'' stars,
particularly in dense areas such as the vicinity of Galactic Plane.
As discussed below (see Sect. 4.4), in a few cases it was necessary
to slightly relax this constraint.

%After numerical simulations the following conditions were adopted:

%Let $\Delta I$ be the magnitude difference in the $I$ band between the
%candidate reference star under selection, and a disturbing neighbour, 
%and $\rho$ the angular separation in arc-sec.
%A star can be retained as reference star only if all neighbours detected
%around verify one of the 3 cases listed below:
%
%
%\begin{itemize}
% \item $\Delta I < 3$: $\rho> 80$ arc-sec
% \item $3\leq \Delta I \leq 4$: $\rho> 60$ arc-sec
% \item $\Delta I >4$ : whatever the separation
%\end{itemize}

\subsection{Hipparcos stars}

We also felt that all the RVS standard stars should be in the Hipparcos 
catalogue \citep[see][]{ESA97}.  
This allows the use of only a few common criteria 
over very homogeneous data acquired over the full sky with a unique instrument: 
magnitudes, variability, indications of multiplicity or close neighbours.
The Hipparcos magnitudes and colours $V$ and $V-I$ are also used to 
calculate $G_{RVS}$, even if
the $V-I$ colours are not perfect \citep[see][]{Platais}.

A (small) drawback is the magnitude distribution of the Hipparcos stars, which drops 
strongly beyond V= 9; however the RV catalogues used below contain mostly HIP stars,
so that the real loss due to this requirement is not very important.

\subsection{Hipparcos master list}

A first list was extracted from the Hipparcos Catalogue, taking into
account all the above criteria, except those on neighbourhood and initial RV 
accuracy. It contains some  42109 stars, with no RV indication.
Some newly recognized doubles found in the SIMBAD data base were then removed.

Then the problem of neighbourhood was handled with 
use of USNO-B1 catalogue, which is available on-line at CDS Strasbourg; 
it contains an $I$ magnitude for most objects.
Each star of the above list was examined in the 
USNO-B1 catalogue, and all its neighbours within 80 arc-sec and brighter
than $I=15$ were retrieved. Because for the faintest stars an $I$ magnitude
is not always available, but only $B$ and $R$, it was roughly estimated by:
$I= R - 0.34\times  (B-R)$  (empirical relation established with the bright stars). 
% also faked faint stars were eliminated as much as possible by various conditions. 
Of course this treatment is not totally
rigorous; it was however considered as the best one, owing to the
complexity of the problem. If some stars are not really good, the
ongoing re-observation programme should help eliminate most
of them. It should be noted that the HIP stars eliminated through
this ``neighbourhood condition'' are of course more numerous in
dense areas, particularly in the vicinity of the Galactic Plane.

The choice of USNO-B1 \citep[see][]{USNOB} catalogue was motivated 
by the following considerations:
full-sky coverage; very large number of stars; very large magnitude range, 
with the bright stars being taken from the Tycho-2 catalogue; 
availability of $I$ magnitude for most stars handled ($I \leq15$); 
good access to the on-line catalogue at CDS Strasbourg 
through the Vizier interface.
A recent partial check on southern selected stars based on 
the newly-available UCAC3 \citep[see][]{UCAC3} 
showed that only a very small fraction of retained 
stars still had disturbing neighbours not listed in USNO-B1; 
however the UCAC3 has a smaller
magnitude range and lacks $I$ magnitudes for most of the stars.

All stars satisfying the above neighbourhood conditions were kept.
The resulting list of 38\,169 HIP stars is the ``master list'': only
stars within it can be accepted. 
Stars selected afterwards from RV criteria will be considered as usable
only if they are in this master list. However, as will be shown below
(see Sect. 4.4), it will be necessary in a few cases to take
stars not found in the master list because of too many disturbing
neighbours.

\section{Selection among existing published RV catalogues}

Several recently published catalogues of reliable RV have been examined.
The aim is to cover the sky roughly homogeneously with accurate 
data coming from only a \textit{small }
number of spectrographs, which have some stars in common. Therefore, although
many recent catalogues are available in the literature, only three were retained 
beside the official IAU standards:
those of \citet{Nidever}, \citet{Nordstrom}, and \citet{Famaey}.
All have different presentations, qualities and parameters; and the selection
must be adapted to each one. The intersection with our master list allows us to 
concentrate on only RV criteria; the others are automatically fulfilled.

For reading and citation convenience, these catalogues will be denoted below: 
NID, NOR, FAM, in addition to IAU.

\subsection{The Nidever et al. (2002) catalogue = NID}

This catalogue contains 889 stars observed during four years between 1997 and
2001, and 782 exhibit velocity scatter smaller than 0.1\,km s$^{-1}$.
Nonetheless, ``...they suffer
from three sources of systematic errors, namely convective blueshift, 
gravitational redshift, and spectral mismatch of the reference spectrum...
The spectra were obtained with the HIRES spectrometer on the 10m Keck-1
telescope, and with the Hamilton echelle spectrometer fed either by the
3-m Shane or the 0.6 CAT...'' (as taken from the Catalogue's Read-Me at
CDS, Vizier Section). Calibration was done by iodine vapour lines
superimposed over the stellar lines. Each programme star has an average
of 12 observations.

Only the HIP stars of their Table 1 have been taken into account (stable stars 
with rms $\leq$100\,m s$^{-1}$, 742 rows). Of them, only 329 are in our
``master list'', mainly due to the requirement of neighbourhood.

We note that the ``three sources of systematic errors'' mentioned above 
may apply as well to the other catalogues.

\subsection{The Nordstr\"{o}m et al. (2004) catalogue = NOR}

This catalogue (``The Geneva-Copenhagen survey of the solar neighbourhood'') 
contains a complete, magnitude-limited,
and kinematically unbiased sample of 16\,682 nearby F and G dwarf stars.
Complete kinematic information for 14\,139 stars is taken from Hipparcos/
Tycho data, and complemented by new, accurate radial-velocity observations 
for most of the stars. The aim was to obtain new  determinations of 
metallicity, rotation, age, kinematics, and Galactic orbits for this sample,
and new  isochrone  ages  for  all  stars  for  which  this  is  possible.
The number of RV observations used and the corresponding time span are 
available for each star.

From Table 1 of the NOR  catalogue
we extracted the HIP stars with at least 2 
RV observations, mean error and standard deviation
on RV smaller than 0.3\,km s$^{-1}$, and considered as single by the authors:
3227 stars before intersection with our ``master list'', all with RVs
obtained with CORAVEL, north and south.
1696 stars have only two measurements, and 1531 have at least three.
1039 stars with at least three measurements are in the master list.

\subsection{The Famaey et al  (2005) catalogue = FAM}

This catalogue provides Hipparcos positions, Hipparcos and Tycho-2 proper
    motions, and CORAVEL radial velocities for 6691 K and M giants in the
    solar neighbourhood, mostly from the Hipparcos survey (northern 
hemisphere only).
It is intended for deriving
    the kinematics of giant stars in the solar neighbourhood, and to
    correlate it with their location in the Hertzsprung-Russell diagram.
Binaries for which no centre-of-mass velocity could be estimated have been 
excluded; but known binaries remain, and will have to be eliminated.
The primary sample includes 5952 K giants and 739 M giants. 86\% are
in the Hipparcos ``Survey'' (for the definition of the Hipparcos Survey,
see \citet{Crifo}.
Useful complementary data on the radial velocities, not available in the
published tables (number of measurements, time span between first 
and last observation) were sent very kindly by B. Famaey.
As most of these stars have only a small number of RV measurements, 
the Bizyaev criterion \citep{Bizyaev} was applied in 
order to increase the probability of retaining 
stable giants; with the available colours at our
disposal this criterion may be converted into:
$0.9 < B-V < 1.2$.  
Only the FAM stars are concerned: the NID stars have
many more observations, and only ``stable'' stars have been selected;
the NOR stars include no giants.
Before intersection with our ``master list'', 1586 FAM stars satisfied
this criterion (K giants only).
Of them, 1449 stars have only two measurements, and 137 have at least three.
126 stars with at least three measurements are in the master list.

\subsection{Resulting list}

In a first step, only single stars of the 4 RV-catalogues (IAU, NID, 
NOR, FAM) with an RV accuracy better than 0.3\,km s$^{-1}$,
and at least three existing measurements within 
the NOR and FAM catalogues were kept and cross-matched with 
our ``master list'';
the result is a list of 1342 stars, a sufficiently large number of objects
for our purpose; but some regions of the sky were not very well covered,
particularly the vicinity of the galactic plane. The selection criteria for 
neighbourhood  were
therefore slightly relaxed (vicinity distance down to 60$\arcsec$ instead 
of 80$\arcsec$ for $3\leq \Delta I \leq 4$, after numerical simulations);
also in some places, stars with only two existing RV measurements  from
NOR or FAM had to be taken in order to fill the gaps. 
This supplementary list  (not all in the the master list) was selected by hand 
and contains only 78 objects. Of course these stars must be re-observed 
in priority, with two new measurements before Gaia launch.

The current list of candidate standard stars contains 1420 stars.
Some stars are common to two or even three RV lists and are therefore 
particularly useful.
Notice that some stars might be rejected in the future, 
after more remeasurements are performed.

\section{Statistical properties of the list}

\subsection{Sky distribution}

\begin{figure}[htp]

\begin{center}
%\resizebox{\hsize}{!}{\includegraphics[angle=-90]  {figures/carteg8liste.ps}}
\resizebox{\hsize}{!}{\includegraphics[angle=-90]  {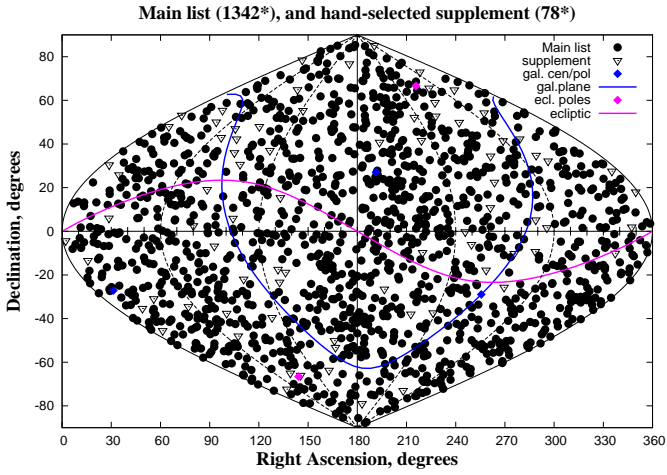}}

 \caption{Map of the selected stars: full circles are the 1342 best stars; triangles
are the 78 slightly less good ones selected by hand in the gaps, mostly near the 
galactic plane }
 \label{carteg8liste}
\end{center}
\end{figure}

The full sky distribution for these 1420 stars is shown in Fig. 
\ref{carteg8liste}, with different symbols for the 1342 best and 78 
supplementary stars. The gaps in the vicinity of galactic plane
are not perfectly filled.

\begin{figure}[htp]

\begin{center}
%\resizebox{\hsize}{!}{\includegraphics[angle=-90]  {figures/carteg8origines.ps}}
\resizebox{\hsize}{!}{\includegraphics[angle=-90]  {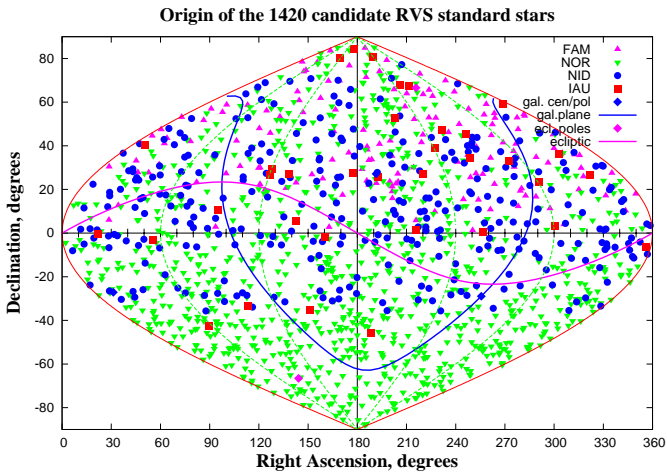}}
 \caption{Origin of the selected stars per list. Most of them are from 
NOR, especially in the south, thanks to the southern CORAVEL.}
 \label{carteg8origines}
\end{center}
\end{figure}

Figure \ref{carteg8origines} shows the origin of the selected stars,
according to the following hierarchical order for stars found in various lists:
IAU; NID; NOR; FAM. 
Stars found in several lists are plotted according to their highest-priority list. 

\subsection{Distribution in magnitude}

\begin{figure}[htp]

\begin{center}
%\resizebox{\hsize}{!}{\includegraphics[angle=-90]  {figures/g8histovmag.ps}}
\resizebox{\hsize}{!}{\includegraphics[angle=-90]  {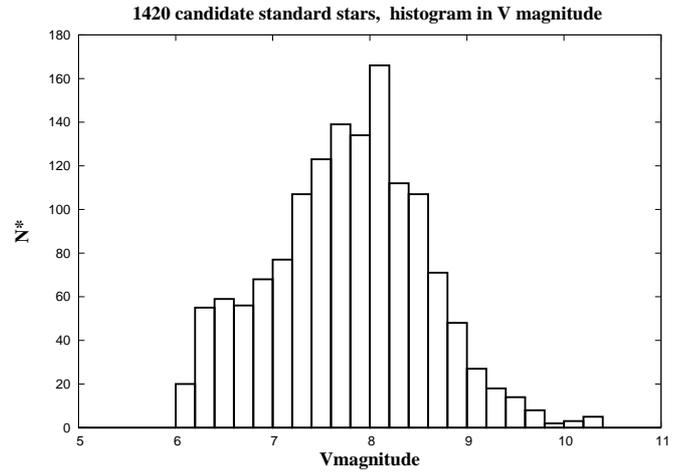}}

  \caption{The list of 1420 candidates: histogram in $V$ magnitude }
 \label{g8histovmag}
\end{center}
\end{figure}

Figure \ref{g8histovmag} shows the histogram in $V$ magnitude: 
its mode is at about 8.1, slightly lower than in Hipparcos 
(8.6); this difference is due mainly to the important effort in the Hipparcos
catalogue for \textit{reducing} the total number of cool stars and particularly
the red giants included in the so-called `` Hipparcos survey'' (see \citet{Crifo});
and here we take only cool HIP stars.

\subsection{Distribution in colour}

\begin{figure}[htp]
\begin{center}
%\resizebox{\hsize}{!}{\includegraphics[angle=-90]  {figures/hrbv.ps}}
\resizebox{\hsize}{!}{\includegraphics[angle=-90]  {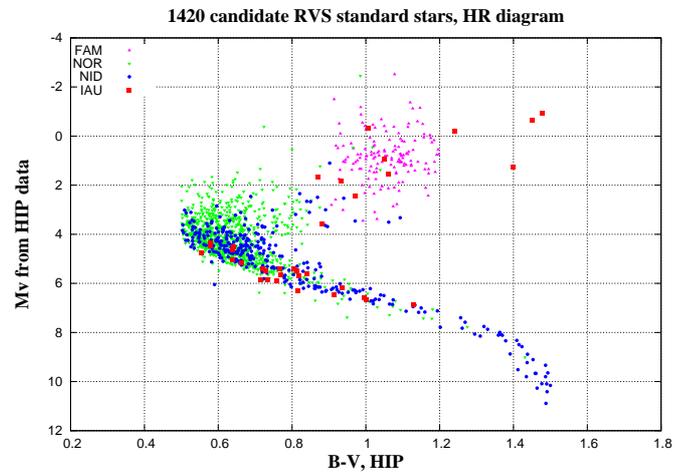}}

 \caption{HR diagram of the stars. Colour code and hierarchy for origin as
in Figure \ref{carteg8origines}}.
\label{hrbv}
\end{center}
\end{figure}

Figure \ref{hrbv} shows the HR diagram of the stars, with the same colour code as 
for the map of Fig. \ref{carteg8origines}. 
The separation between the various origins is very clear.
Most giants are from FAM, and therefore occupy  an area limited
 by the Bizyaev criterion.
 
\subsection{Distribution in Radial Velocities}

\begin{figure}[htp]

\begin{center}
%\resizebox{\hsize}{!}{\includegraphics[angle=-90]  {figures/virv.ps}}
\resizebox{\hsize}{!}{\includegraphics[angle=-90]  {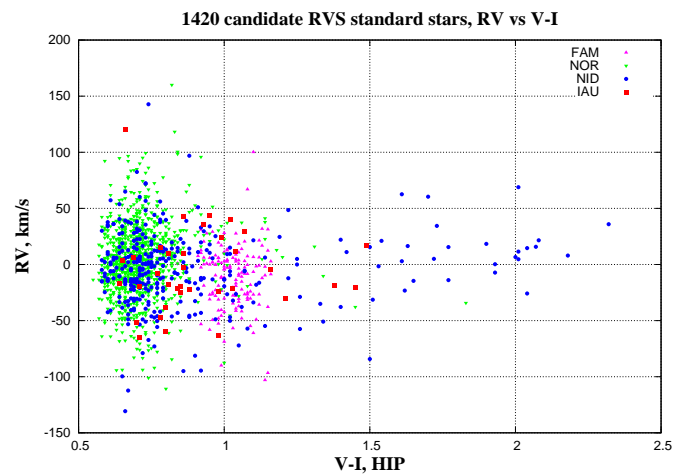}}

 \caption{Distribution in radial velocities.
Colour code and hierarchy as in Figure \ref{carteg8origines} }
\label{virv}
\end{center}
\end{figure}

Figure \ref{virv} shows the distribution in radial velocities vs 
$(V-I)_{HIP}$. 
The highest RVs come from NOR; but are
nevertheless not extremely high (only one star above 150\,km s$^{-1}$).

\section{Comparison between the different lists}

Our list is made out of four previously existing lists. These have several 
stars in common, but with data obtained with  different instruments.

Table 1 shows the numbers of common stars. It also gives the number of stars
retained within each original catalogue. These numbers are slightly larger
than those given for the intersection with the master list, due to the
supplementary list of 78 stars.

\begin{table}[!h]
\begin{center}
%\begin{tabular}{|c|cccc|}
\begin{tabular}{c|cccc}
\hline\hline
 & IAU & NID & NOR & FAM\\
\hline
IAU & 35 & 16 & 16 & 3 \\
NID & 16 & 336 & 163 & 3 \\
NOR & 16 & 163 & 1084 & 1\\
FAM & 3  & 3   & 1    & 154 \\
\hline
\end{tabular}
\caption{Numbers of stars  common to various lists}
\end{center}
\end{table}

\begin{figure}[htp]

\begin{center}
%\resizebox{\hsize}{!}{\includegraphics[angle=-90]  {figures/difuai.ps}}
\resizebox{\hsize}{!}{\includegraphics[angle=-90]  {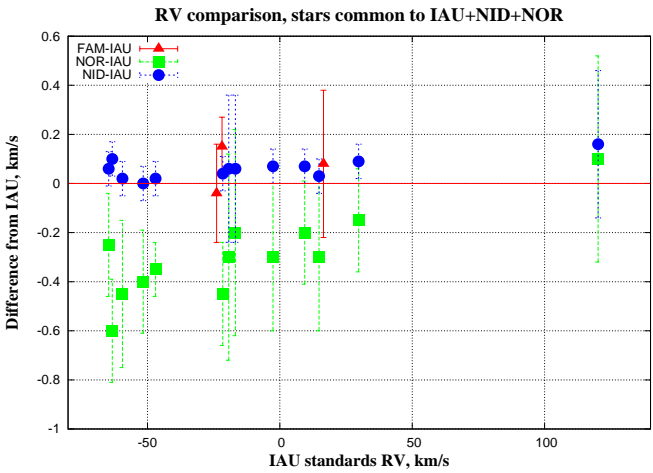}}

 \caption{The 13 stars common to IAU, NID and NOR ; plus the 3 IAU/FAM stars,
all compared to IAU RV.  } 
\label{difuai}
\end{center}
\end{figure}

It is interesting to compare the data of the common stars.
IAU standards are of course taken as references; but they have quite 
 small intersections with other lists. Only 13 stars are common to 
IAU, NID and NOR ; and the FAM stars contain only 3 IAU stars, not 
found in the other lists.
Figure \ref{difuai} shows these 13 common stars, plus the 3 FAM stars.

 The offset between NID+FAM and IAU is quite small (0.063\,km s$^{-1}$); but the
NOR data for the same objects show a larger offset, and with 
larger error bars due to the less accurate CORAVEL measurements.

\begin{figure}[htp]	%fanornidif

\begin{center}
%\resizebox{\hsize}{!}{\includegraphics[angle=-90]  {figures/fanornidif.ps}}
\resizebox{\hsize}{!}{\includegraphics[angle=-90]  {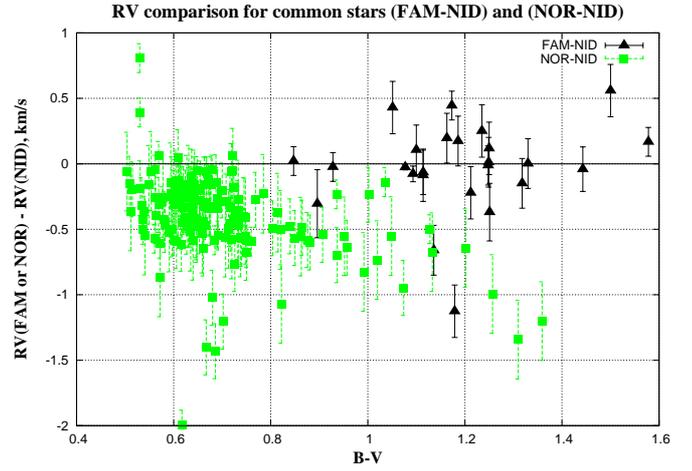}}

\caption{ Comparison of NOR and FAM with NID, vs $B-V$.
 The 163 stars common to NOR and NID
are those of Table 1, and the 25 stars common to FAM and NID are 
taken before intersection with the master list.}

\label{fanornidif}
\end{center}
\end{figure}

Because the samples involved in Fig. \ref{difuai} are very small, 
we investigated  the differences between the NOR, FAM and NID data more
in detail, taking now the NID stars as references.
For NOR vs NID we took the 163 common objects of Table 1.
For FAM vs NID:
as the number of common stars is again very small (only 3, see Table 1),
we tried to increase this number by comparing the lists \textit{before}
their intersection with our master list. A set of 25 common stars could be 
defined that way, including the 3 ones of Table 1. 
In Fig. \ref{fanornidif} the differences in RV (FAM - NID) and
(NOR - NID) are plotted versus $B-V$. Although the FAM
stars cover only a small interval in $B-V$, they clearly  
agree well with NID, while the NOR data show a trend 
in $B-V$.

These results are derived from the \textit{published} data. However, the 
NOR and FAM data are for a good part obtained with the same
instrument (northern CORAVEL, located at OHP, France), and are handled
in the same Coravel database.
It appears that the NOR data, which were obtained and published before the 
FAM ones, experienced an earlier reduction in the database; 
and a small change in the scale was introduced between the two sets.
The recent updates of the NOR catalogue by Holmberg et 
al \citep{Holmberg2007, Holmberg2009}
did not update the radial velocities.
Clearly, the new scale used for the FAM data allows a much
better agreement with the NID data.
For the final RVS list, we will use the NOR data re-reduced with the 
new scale.

The offsets seen in Fig. \ref{difuai} and \ref{fanornidif} 
reveal the differences of RVZP proper 
to each instrument and reduction procedure related to 
resolution, wavelength ranges, calibration lamps, line lists, numerical 
tools for fitting the lines, template/masks (models) used for the correlation 
function (K0, G5, G2) etc... The aim of the observing programme that we have 
started to follow-up the list of candidates is not only to check the RV 
stability of the stars, but also to define a common RV scale based on 
asteroids (see next section). As a consequence our final catalogue of RV 
standards will be homogeneous, with an absolute RVZP.

%%%%%%%%
%\input{section_obs_modifFC100714.tex}

\section{Re-observation programme}   	% version de FC du 14/7/2010

From the list of 1420 candidates, at least 1000 RV-STD stars will be selected. 
These primary stars have to be stable in radial velocity at the 300\,m s$^{-1}$
 level,
with no drift until the end of the mission (2018). To be qualified as a 
RVS standard star, each candidate has to be observed at least  twice until the 
late mission phase in order to verify its long term stability.

Our observational strategy has been defined to respond to the following 
constraints:

\begin{itemize}
\item Checking the 300\,m s$^{-1}$ stability implies a RV accuracy
of individual measurements  better than 100\,m s$^{-1}$, which is easily 
obtained with modern echelle spectrographs and FGK stars, without 
requiring a high signal-to-noise ratio. 
\item To cover the whole sky, we need to use northern and southern instruments, 
which implies  checking the consistency between the
different instruments and to set the zero-point scale. Asteroids and IAU 
standards are systematically observed for that purpose.
\item The NID list  is supposed to be already cleaned of non-stable 
stars at the level of 100\,m s$^{-1}$.
 Only one supplementary observation is needed before launch to verify 
the stability of NID candidates.
\item FAM and NOR catalogues rely on CORAVEL measurements, 
which have a typical uncertainty of 300\,m s$^{-1}$.
Moreover most candidates have only three or even two RV measurements available. 
At least two supplementary observations
are mandatory to check their stability before launch.
\item For all non-rejected candidates at the time of launch, one more 
observation during the mission has to be done to reject stars with
long-term variations. 
\end{itemize}

To conduct this long-term and extensive programme, three echelle 
spectrographs are used: SOPHIE mounted on  the T193 at OHP,
NARVAL mounted on the Telescope Bernard Lyot at Pic du Midi Observatory; 
and CORALIE mounted on the Swiss Euler Telescope at La Silla. 

We have also collected  useful measurements in the ELODIE 
archive \citep{ELODIE}, and will search the HARPS archive at ESO.

All these instruments provide RV measurements with an accuracy better 
than 50\,m s$^{-1}$.

The re-observation programme is going on and should be finished for its
main part around mid-2012; then the data have to be compared, and
homogeneized between all telescopes, thanks to the use of IAU standards,
the stars common to several telescopes, and asteroids.
Therefore the observing programme includes the observation of one to four 
or five asteroids per night. 
The zero-point adopted for the RV scale will be the spectroscopic RV of
asteroids; it is correlated with kinematic radial velocities
given by the  ephemerides, wich have a typical error of 1\,m s$^{-1}$.
However this correlation depends on computed convective shifts
in the solar spectrum  and physical properties of the asteroids, such as
angular diameter, spin, phase... This point will be developed later 
at the end of the observations.

\section{Availability of the list}

As already stated in several paragraphs above, the present list
is only \textit{preliminary}; 
the version to be used during the mission, with probably less objects
but new RVs, will be ready for Gaia launch;
the final version will be known only at the end of the Gaia mission, around 2018.
The present preliminary list of 1420 objects is available at CDS, in the 
Catalogues section. It contains only already published data, extracted
from the various original catalogues described above.
A subset is given in Table 2. Only the data needed for a good identification
are kept, plus the radial velocity data from the various original sources, 
and the way it was introduced here: the main list of 1342 stars, or the 
supplementary list of 78 stars.

\section{Conclusion}
A new preliminary full-sky list of candidate Radial Velocity standard stars has been established,
containing 1420 HIP stars with a present accuracy of 300\,m s$^{-1}$ or better at selection, 
to be improved to 100 or possibly 50\,m s$^{-1}$ after re-observation. 
The magnitude range of these cool stars is 6 to 11 in V. 
The stars are well isolated (no disturbing neighbours within 80$\arcsec$, or 60$\arcsec$
in a few cases in dense areas).
An important reobservation programme is going on, because the stability in RV has to be
guaranteed until the end of the Gaia mission (about 2018).
Compared to the existing IAU RV-standards, it is a denser and fainter extension,
with a good full-sky coverage.
It is designed primarily for the calibration of the Gaia-RVS, but 
may be used for many other purposes and projects.

%\section{Ce qui manque encore}

%===================================================

\begin{acknowledgements}

% CU6 et DPAC:
 Apart from the authors, this work was encouraged and supported by all the 
RVS team, within the Gaia DPAC Consortium (Data Processing  and  Analysis  
Consortium, led by F. Mignard); and by the french ``Action Sp\'ecifique Gaia''
led by C. Turon.

% CDS:
We warmly thank the staff of CDS-Strasbourg, where most of the data are found,
for the SIMBAD, Aladin and Vizier tools, for their so friendly help on all 
types of subjects, and particularly F. Ochsenbein.

% Famaey pour ses donnees supplementaires
It is also a pleasure to thank Dr B. Famaey, for providing us so kindly with
additional unpublished data that are very useful for the selection process.

% Nos PN et Geneve, les observateurs:
Also, this work would be almost useless without the ground-based re-observing 
programme: the many northern telescope nights at Pic du Midi on Narval,
and at Observatoire de Haute Provence (OHP) on SOPHIE are taken in charge by 
the french National Programmes PNPS and PNCG, with often queue observations 
made by the local staff. 
The southern observing nights have been carried out with the 
CORALIE spectrograph mounted on the swiss Euler Telescope at ESO, 
La Silla (Chile).

% Autres: nos obs? Nos labos?
Our thanks also go the referee, whose sharp eye detected many inconveniences,
and insufficient explanations and justifications.

\end{acknowledgements}

%===================================================

% La biblio par bibtex, dans le fichier biblio.bib
%\bibliographystyle{aa}          % style aa.bst
%\bibliography{biblio}           % fichier de biblio biblio.bib separe

%\begin{sidewaystable}
%\centering
\begin{table*}
%\begin{center}
%\hspace{-3cm} \hspace{0.5cm}
\begin{tabular}{|rl|ll|rrrl|cr|rr|rrr|rrr|c|} \hline\hline
 HIP & HD/DM & \multicolumn{2}{c|}{coordinates, J(2000, 2000) } & V  &  B-V   & V-I  &   Sp.T       &
\multicolumn{2}{c|}{IAU} & \multicolumn{2}{c|}{NID} & \multicolumn{3}{c|}{NOR } & \multicolumn{3}{c|}{FAM} & List \\     
    &       &                     &                           &    &        &      &              &  E/C &   RV             &  RV & $\Delta T$            &  RV   & N  & $\Delta T$        & RV   & N  & $\Delta T$ &    \\
    &       &                     &                           &    &        &      &              &      &  km s$^{-1}$
& km s$^{-1}$ & days                  & km s$^{-1}$  &    &  days             & km s$^{-1}$ &    & days       &     \\
%  HIP &  HD        &   alfa       &  delta        &  V    &   B-V  & V-I  &    SpT          & IAU1     &   IAU2     & nid1     &  nid2    &  nor1    & nor2     & nor3     &  fam1    & fam2     & fam3     & L \\
\hline 
   296 & HD 225118  & 00:03:41.484 & -28:23:46.304 &  8.24 &  0.780 & 0.82 &    G8.5V        & $\cdots$ &   $\cdots$ & $\cdots$ & $\cdots$ &   10.6   &   4      & 3656     & $\cdots$ & $\cdots$ & $\cdots$ & M \\
   407 & HD 225299  & 00:04:58.716 & -70:12:44.802 &  8.13 &  0.710 & 0.76 &    G5V          & $\cdots$ &   $\cdots$ & $\cdots$ & $\cdots$ &   12.1   &   4      & 2170     & $\cdots$ & $\cdots$ & $\cdots$ & M \\
   420 & HD 23      & 00:05:07.495 & -52:09:06.268 &  7.53 &  0.577 & 0.65 &    G0V          & $\cdots$ &   $\cdots$ & $\cdots$ & $\cdots$ &   34.3   &   3      & 1767     & $\cdots$ & $\cdots$ & $\cdots$ & M \\
   466 & HD 58      & 00:05:34.930 & +53:10:18.097 &  7.23 &  1.174 & 1.14 &    K0           & $\cdots$ &   $\cdots$ & $\cdots$ & $\cdots$ & $\cdots$ & $\cdots$ & $\cdots$ &   -9.72  &   2      &  812     & S \\
   556 & HD 200     & 00:06:46.972 & -04:21:00.095 &  8.21 &  0.574 & 0.65 &    F8           & $\cdots$ &   $\cdots$ & $\cdots$ & $\cdots$ &   -0.2   &   2      &  360     & $\cdots$ & $\cdots$ & $\cdots$ & S \\
   616 & HD 283     & 00:07:32.541 & -23:49:07.390 &  8.70 &  0.798 & 0.84 &    G9.5V        & $\cdots$ &   $\cdots$ &  -43.102 & 1185     & $\cdots$ & $\cdots$ & $\cdots$ & $\cdots$ & $\cdots$ & $\cdots$ & M \\
   624 & HD 307     & 00:07:37.430 & -45:07:10.337 &  8.19 &  0.584 & 0.66 &    F9V          & $\cdots$ &   $\cdots$ & $\cdots$ & $\cdots$ &   14.6   &   4      & 1477     & $\cdots$ & $\cdots$ & $\cdots$ & M \\
   699 & HD 400     & 00:08:40.938 & +36:37:37.650 &  6.21 &  0.504 & 0.58 &    F8IV         & $\cdots$ &   $\cdots$ &  -15.141 &  507     &  -15.2   &  20      & 4858     & $\cdots$ & $\cdots$ & $\cdots$ & M \\
   726 & HD 457     & 00:08:59.678 & -39:44:13.773 &  7.72 &  0.620 & 0.69 &    G0V          & $\cdots$ &   $\cdots$ & $\cdots$ & $\cdots$ &  -19.4   &   3      & 1477     & $\cdots$ & $\cdots$ & $\cdots$ & M \\
   791 & HD 547     & 00:09:48.559 & -40:53:34.562 &  8.57 &  0.644 & 0.71 &    G3V          & $\cdots$ &   $\cdots$ & $\cdots$ & $\cdots$ &   14.3   &   3      & 1476     & $\cdots$ & $\cdots$ & $\cdots$ & M \\
   801 & HD 564     & 00:09:52.821 & -50:16:04.161 &  8.32 &  0.595 & 0.67 &    G2/G3V       & $\cdots$ &   $\cdots$ & $\cdots$ & $\cdots$ &   11.1   &   5      & 3659     & $\cdots$ & $\cdots$ & $\cdots$ & M \\
   867 & HD 631     & 00:10:39.127 & +12:49:12.830 &  8.47 &  0.574 & 0.65 &    G5           & $\cdots$ &   $\cdots$ & $\cdots$ & $\cdots$ &   32.3   &   4      & 1162     & $\cdots$ & $\cdots$ & $\cdots$ & M \\
   980 & HD 763     & 00:12:06.826 & +47:29:27.964 &  7.36 &  1.082 & 1.04 &    K0           & $\cdots$ &   $\cdots$ & $\cdots$ & $\cdots$ & $\cdots$ & $\cdots$ & $\cdots$ &    4.36  &   4      &  369     & M \\
  1031 & HD 870     & 00:12:50.250 & -57:54:45.394 &  7.22 &  0.775 & 0.82 &    K0V          & $\cdots$ &   $\cdots$ & $\cdots$ & $\cdots$ &    1.0   &   4      & 2248     & $\cdots$ & $\cdots$ & $\cdots$ & M \\
  1147 & HD 1000    & 00:14:20.800 & -21:11:49.063 &  6.89 &  0.502 & 0.57 &    F7V          & $\cdots$ &   $\cdots$ & $\cdots$ & $\cdots$ &  -13.6   &   3      &  925     & $\cdots$ & $\cdots$ & $\cdots$ & M \\
\hline
\end{tabular}
\caption{
Extract of the preliminary list of new RV standards, available at CDS in electronic form.
See note(1) for a description of the column contents. 
}    % fin de la caption
%\footnotemark{}
 
Contents of  Table:
   Column (HIP, HD/DM): identifiers.
  Col. (coordinates): all in J(2000, 2000).
  Col. (V, B-V...): V, B-V, V-I from HIP; SpT from SIMBAD.
  Col. (IAU): RV from IAU Comm30 standards. E/C= ELODIE or CORAVEL standards.
No IAU std is found in this short extract.
   Col. (NID, NOR, FAM): RV in km s$^{-1}$; N = number of observations used for RV;
$\Delta T$ = time span (days) between first and last observation.
  Col. (List): M indicates star in main list; S in supplementary 
list, i.e. with reduced environment conditions, or with only 2 measurements
 (see Sect. 4.4).
 The two ``S'' stars shown here have only 2 measurements.
 
\end{table*}
%\end{sidewaystable}

\end{document}